\documentclass[11pt, draftcls, onecolumn,journal]{IEEEtran}
\def\pterror#1{\errmessage{Parsetree ERROR: #1}}
\newdimen\pthgap\def\pthorgap#1{\pthgap=#1}
\newdimen\ptvgap\def\ptvergap#1{\ptvgap=#1}
\newbox\ptnodestrutbox\def\ptnodestrut{\unhcopy\ptnodestrutbox}
\newbox\ptleafstrutbox\def\ptleafstrut{\unhcopy\ptleafstrutbox}
\def\ptnodefont#1#2#3{\def\ptnodefn{#1}
  \setbox\ptnodestrutbox=\hbox{\vrule height#2 width0pt depth#3}}
\def\ptleaffont#1#2#3{\def\ptleaffn{#1}
  \setbox\ptleafstrutbox=\hbox{\vrule height#2 width0pt depth#3}}
\pthorgap{6pt}                         
\ptvergap{6pt}                         
\ptnodefont{\normalsize\rm}{9pt}{3pt}  
\ptleaffont{\normalsize\it}{9pt}{3pt}  
\newcount\ptdepth           
\newcount\ptn               
\newbox\ptm \newdimen\ptmx  
\newbox\pta \newdimen\ptax  
\newbox\ptb \newdimen\ptbx  
\newbox\ptc \newdimen\ptcx  
\newbox\ptx \newdimen\ptxx  
\newif\ifpttri              
\def\ptnext{\advance\ptn by 1 \ifcase\ptn
  \or \setbox\ptm=\box\ptx \ptmx=\ptxx \or \setbox\pta=\box\ptx \ptax=\ptxx
  \or \setbox\ptb=\box\ptx \ptbx=\ptxx \or \setbox\ptc=\box\ptx \ptcx=\ptxx
  \else \pterror{More than 3 daughters in (sub)tree}\fi}
\def\ptbegtree{\ptdepth=0}
\def\ptendtree
  {\ifnum\ptdepth>0 \pterror{Mismatched bracketing: too few ')'s!}\fi}
\def\ptbeg{\ifnum\ptdepth=0 \leavevmode\fi\begingroup
  \advance\ptdepth1 \ptn=0\pttrifalse}
\def\ptend{\ifnum\ptdepth=0 \pterror{Mismatched bracketing: too many ')'s!}
  \else\ptcons\endgroup\ifnum\ptdepth=0 \box\ptx\else\ptnext\fi\fi}
\def\ptnodeaux#1{\setbox\ptx=\hbox{#1}\ptxx=0.5\wd\ptx\ptnext}
\def\ptnode#1{\ptnodeaux{\ptnodefn\ptnodestrut #1}}
\def\ptleaf#1{\ptnodeaux{\ptleaffn\ptleafstrut #1}}
\def\pthoradjust#1{\ifcase\ptn
  \or \pthadjbox{\ptm}{#1} \or \pthadjbox{\pta}{#1}
  \or \pthadjbox{\ptb}{#1} \or \pthadjbox{\ptc}{#1}
  \else \pterror{More than 3 daughters in (sub)tree}\fi}
\def\pthadjbox#1#2{\setbox#1=\hbox{\box#1\kern#2}}
\def\ptcons
 {\ifnum\ptn=0 \ptconsz 
  \else
    \ifpttri\ptconstri\else
      \ifcase\ptn \or\ptconsm\or\ptconsma\or\ptconsmab\or\ptconsmabc \fi \fi
    \ptax=\ptxx \advance\ptax-\ptmx \ptbx=0pt
    \ifdim\ptax<0pt \ptbx=-\ptax\ptax=0pt\ptxx=\ptmx \fi
    \setbox\ptx=\vtop{\hbox{\kern\ptax\box\ptm}\nointerlineskip
                      \hbox{\kern\ptbx\box\ptx}}\fi
  \global\ptxx=\ptxx\global\setbox\ptx=\box\ptx}
\def\ptavg#1#2#3{#1=#2\advance#1#3#1=0.5#1}     
\def\ptadv#1#2{\advance#1#2\advance#1\pthgap}   
\def\ptconsz{\ptxx=0pt \setbox\ptx=\vtop{}}     
\def\ptconsm{\ptxx=0pt 
  \setbox\ptx=\hbox{\ptedge{1}{0}{}{}}}         
\def\ptconsma                                   
 {\ptxx=\ptax\setbox\ptx=\vtop{
    \hbox{\ptedge{1}\ptax{}{}}\nointerlineskip
    \hbox{\box\pta}}}
\def\ptconsmab                                  
 {\ptadv\ptbx{\wd\pta}\ptavg\ptxx\ptax\ptbx
  \setbox\ptx=\vtop{
    \hbox{\ptedge{2}\ptax\ptbx{}}\nointerlineskip
    \hbox{\box\pta\kern\pthgap\box\ptb}}}
\def\ptconsmabc                                 
 {\ptadv\ptbx{\wd\pta}\ptadv\ptcx{\wd\pta}%
  \ptadv\ptcx{\wd\ptb}\ptavg\ptxx\ptax\ptcx
  \setbox\ptx=\vtop{
    \hbox{\ptedge{3}\ptax\ptbx\ptcx}\nointerlineskip
    \hbox{\box\pta\kern\pthgap\box\ptb\kern\pthgap\box\ptc}}}
\def\ptconstri                                  
 {\ifcase\ptn\or\setbox\pta\hbox{\kern2\pthgap}\or
  \or\setbox\pta=\hbox{\box\pta\kern\pthgap\box\ptb}
  \or\setbox\pta=\hbox{\box\pta\kern\pthgap\box\ptb\kern\pthgap\box\ptc}\fi
  \ptxx=0.5\wd\pta
  \setbox\ptx=\vtop{
    \hbox{\ptedge{0}{0}{\wd\pta}{}}\nointerlineskip
    \box\pta}}
\newcount\pted          
\newcount\ptedm         
\newcount\pteda         
\newcount\ptedb         
\newcount\ptedc         
\newcount\ptedl         
\newcount\ptedh         
\newcount\ptedhs        
\newcount\ptedvs        
\newcount\ptedtemp      
\def\ptedge#1#2#3#4{\pted=#1%
  \pteda=#2\ifcase\pted\ptedb=#3\or\or\ptedb=#3\or\ptedb=#3\ptedc=#4\fi
  \ptedm=\pteda\advance\ptedm\ifcase\pted\ptedb\or\pteda\or\ptedb\or\ptedc\fi
  \divide\ptedm by 2
  \ptedh=\ptvgap\ptedtemp=\ptedm\advance\ptedtemp-\pteda\divide\ptedtemp by 6
  \ifnum\ptedh<\ptedtemp\ptedh=\ptedtemp\fi
  \unitlength=1sp%
  \begin{picture}(0,\ptedh)
    \ifnum\pted=3 \ptedput\ptedc\fi
    \ifnum\pted=1 \else\ptedput\ptedb\fi
    \ptedput\pteda
    \ifnum\pted=0 \ptedbot\fi 
  \end{picture}}
\def\ptedput#1{\ptedl=#1\advance\ptedl-\ptedm
  \ifnum\ptedl>0 \ptedslope\else
    \ptedl=-\ptedl\ptedslope\ptedhs=-\ptedhs\fi
  \ifnum\ptedhs=0 \ptedl=\ptedh\fi
  \put(\ptedm,\ptedh){\line(\ptedhs,-\ptedvs){\ptedl}}}
\def\ptedbot
 {\ptedtemp=\ptedl\multiply\ptedtemp by \ptedvs\divide\ptedtemp by \ptedhs
  \ifnum\ptedtemp>0\ptedtemp=-\ptedtemp\fi \advance\ptedtemp-1
  \advance\ptedtemp\ptedh\multiply\ptedl by 2
  \put(\pteda,\ptedtemp){\line(1,0){\ptedl}}}
\def\ptedslope
 {\ifnum\ptedl>\ptedh\ptedhs=\ptedl\ptedvs=\ptedh
  \else              \ptedvs=\ptedl\ptedhs=\ptedh \fi
  \divide\ptedhs by 60 \divide\ptedvs by \ptedhs
  \ifnum \ptedvs <  5 \ptedvs=0 \ptedhs=1 \else
  \ifnum \ptedvs < 11 \ptedvs=1 \ptedhs=6 \else
  \ifnum \ptedvs < 13 \ptedvs=1 \ptedhs=5 \else
  \ifnum \ptedvs < 17 \ptedvs=1 \ptedhs=4 \else
  \ifnum \ptedvs < 22 \ptedvs=1 \ptedhs=3 \else
  \ifnum \ptedvs < 27 \ptedvs=2 \ptedhs=5 \else
  \ifnum \ptedvs < 33 \ptedvs=1 \ptedhs=2 \else
  \ifnum \ptedvs < 38 \ptedvs=3 \ptedhs=5 \else
  \ifnum \ptedvs < 42 \ptedvs=2 \ptedhs=3 \else
  \ifnum \ptedvs < 46 \ptedvs=3 \ptedhs=4 \else
  \ifnum \ptedvs < 49 \ptedvs=4 \ptedhs=5 \else
  \ifnum \ptedvs < 55 \ptedvs=5 \ptedhs=6 \else
                      \ptedvs=1 \ptedhs=1 
    \fi \fi \fi \fi \fi \fi \fi \fi \fi \fi \fi \fi
  \ifnum \ptedl < \ptedh
    \ptedtemp=\ptedhs \ptedhs=\ptedvs \ptedvs=\ptedtemp \fi}

\def\ptcatcodes
 {\catcode`(=\active \catcode`)=\active
  \catcode`.=\active \catcode``=\active
  \catcode`~=\active}
{\ptcatcodes
\gdef\ptactivechardefs
 {\ptcatcodes
  \def({\ptbeg\ignorespaces}
  \def){\ptend\ignorespaces}
  \def.##1.{\ptnode{##1}\ignorespaces} 
  \def`##1'{\ptleaf{##1}\ignorespaces}
  \def~{\pttritrue\ignorespaces}}
}

\def\wp{\mathbf{W}}
\def\wpf{W}

\def\R{R}
\def\N{N}
\def\NNNN{\! \! \! \!}

\def\r{r}

\def\m{m}
\def\akappa{\omega_{\K}}
\def\Cum{\mathsf{cum}}

\def\Nset{\mathbb{N}}
\def\Zset{\mathbb{Z}}
\def\Rset{\mathbb{R}}
\def\I{\mathrm{I}}

\def\K{\lambda}
\def\S{{\mathsf{S}}}
\def\F{{\mathcal{F}}}

\def\un{{\mbox{\rm 1\hspace{-0.26em}l}}}
\def\sinc{\mathsf{sinc}}

\newtheorem{theorem}{Theorem}

\newtheorem{proposition}{Proposition}
\newtheorem{lemma}{Lemma}
\newtheorem{remark}{Remark}

\newtheorem{corollary}{Corollary}
{\hspace{\stretch{1}}%
\rule{1ex}{1ex}}

\usepackage{graphicx,epsfig}
\usepackage{amssymb}
\usepackage{latexsym}
\usepackage{amsmath}
\usepackage{mathrsfs}
\usepackage{bbm}
\usepackage{balance}

\date{}

\begin{document}


\title{{{Central Limit Theorems for Wavelet Packet \mbox{Decompositions of Stationary Random Processes}}}}

\author{Abdourrahmane M. Atto$^1$ \thanks{$^1$ am.atto@telecom-bretagne.eu,},
\and
Dominique Pastor$^2$\thanks{$^2$ dominique.pastor@telecom-bretagne.eu}
\\
Institut TELECOM, TELECOM Bretagne,
Lab-STICC, CNRS, UMR 3192,\\
Technop\^ole Brest-Iroise,
CS 83818, 
29238 Brest Cedex 3, FRANCE}


\maketitle

\begin{abstract}

This paper provides central limit theorems for the wavelet packet decomposition of stationary band-limited random processes. The asymptotic analysis is performed for the sequences of the wavelet packet coefficients returned at the nodes of any given path of the $M$-band wavelet packet decomposition tree. It is shown that if the input process is centred and strictly stationary, these sequences converge in distribution to white Gaussian processes when the resolution level increases, provided that the decomposition filters satisfy a suitable property of regularity. For any given path, the variance of the limit white Gaussian process directly relates to the value of the input process power spectral density at a specific frequency. 

\end{abstract}



\medskip
\section{Introduction}\label{Intro}

\IEEEPARstart{T}{his} paper addresses the statistical properties of the $M$-Band Discrete Wavelet Packet Transform, hereafter abbreviated as $M$-DWPT.
In \cite{pesquet99}, asymptotic analysis is given for the correlation structure and the distribution of the $M$-Band wavelet packet coefficients of stationary random processes. 
The limit autocorrelation functions and distributions are shown to be the same for every $M$-DWPT path. This seems to be a paradox because the $M$-DWPT paths are characterised by several sequences of wavelet filters. Two arbitrary sequences are different, and thus, do not have the same properties. In addition, the results presented in \cite{pesquet99} seem to be in contradiction with those stated in \cite{atto07} concerning the autocorrelation functions of the standard discrete wavelet packet transform ($M$-DWPT with $M=2$) of wide-sense stationary random processes. The results presented in \cite{atto07} highlight that the limit variance of the wavelet packet coefficients does depend on the path followed in the wavelet packet tree.

In fact, as shown below, the limit results in \cite[Corollary 5, Proposition 12]{pesquet99} apply to only one path of the $M$-DWPT, namely the standard approximation path (only low-pass filters are used in this path). 
The same holds true for \cite[Proposition 7]{pesq07}, which extends \cite[Corollary 5]{pesquet99} to the case of the dual-tree $M$-DWPT. 
Actually, the limit analysis of the autocorrelation and distributions of the $M$-DWPT coefficients is more intricate than presented in \cite{pesquet99} and \cite{pesq07} because, as shown below, this analysis depends on the path chosen and the wavelet filters used for decomposing the input random process. This analysis is presented for the Shannon $M$-DWPT filters and standard families of paraunitary filters that converge to the Shannon paraunitary filters.


\section{Preliminary results}\label{Mband}

\subsection{General formulas on the $M$-DWPT}\label{Mband1}

In what follows, $j$ and $M$ are natural numbers and $M\geqslant 2$. An $M$-DWPT is performed by using wavelet paraunitary filters with impulse responses $h_{\m}, \m =0, 1, 2, \ldots, M-1$. For further details about the computation and the properties of $M$-DWPT filters, the reader is asked to refer to \cite{steffen93}. 
Let
\begin{equation}
H_{\m}(\omega) = \frac{1}{\sqrt{M}}  \sum_{\ell \in \mathbb{Z}} h_{\m}[{\ell}] \exp \left ( {-i\ell \omega} \right ),
\label{eqHm}
\end{equation}
and $\Phi$ be a function such that $\{\tau_k \Phi~: k\in \mathbb{Z}\}$ is an orthonormal system of $L^2(\mathbb{R})$, where $ \tau_k \Phi~: t \longmapsto \Phi(t-k)$. Let $\mathbf{U}$ be the closure of the space spanned by this orthonormal system. 

The $M$-DWPT decomposition of the function space $\mathbf{U}$ involves splitting $\mathbf{U}$ into $M$ orthogonal subspaces (an easy extension \cite[Lemma 10.5.1]{daub92} established for the standard DWPT): 
\begin{equation}
\mathbf{U} = \bigoplus_{\m=0}^{M-1}  \wp_{1,\m},
\label{eq10avanta}
\end{equation}
and recursively applying the following splitting 
\begin{equation}
\wp_{j,n} = \bigoplus_{\m=0}^{M-1}  \wp_{j+1,Mn+\m},
\label{eq10a}
\end{equation}
for every natural number $j$ and every $n = 0, 1, 2, \ldots, M^j-1$. In this decomposition, the \emph{wavelet packet space} $\wp_{j,n}$ is the closure of the space spanned by the orthonormal set of the \emph{wavelet packet functions} $\{ \wpf_{j,n,k} : k \in \mathbb{Z}\} $ whose Fourier transforms are given by 
\begin{equation}
\F{\wpf_{j, n, k}}(\omega)
=
\exp \left ( {-i M^j k \omega} \right ) \F{\wpf_{j, n}}(\omega),
\label{eqWPFjnk}
\end{equation}
with
\begin{equation}
\label{eqWPFwjn}
\F \wpf_{j,n}(\omega) = M^{j/2} \F \wpf_n(M^j \omega),
\end{equation}
where the sequence $\wpf_{n}, n=0,1,2, \ldots,$ is recursively defined by
\begin{equation}
\label{eq winit}
\F \wpf_m(\omega) = H_m( \omega / M ) \F \Phi ( \omega / M ), 
\end{equation}
and
\begin{equation}
\F \wpf_{Mn+\m}(\omega) = H_{\m}(\omega / M) \F \wpf_{n}(\omega / M),
\label{eqWPF-bis}
\end{equation}
for $m = 0, 1, 2, \ldots, M-1$ and $n \in \Nset$.

Note that the function $\Phi$ in Eq. \eqref{eq winit} is not necessarily the standard scaling function associated with the low-pass filter $h_0$ (see \cite[Lemma 10.5.1]{daub92} for more details). If $\Phi$ is this scaling function, we have $\wpf_{0}=\Phi$ in Eq. \eqref{eq winit}.

\subsection{$M$-ary representations of the $M$-DWPT paths}\label{Mband1-}

A given wavelet packet path $\mathcal{P}$ is described by a sequence of nested functional subspaces: $\mathcal{P}=\left(\mathbf{U}, \{\wp_{j,n(j)}\}_{j \in \Nset}\right)$, where $\wp_{j,n(j)}\subset \wp_{j-1,n(j-1)}$, with $n(0) = 0$ by convention and
\begin{equation}
\label{eqdpter}
n(j) = M n(j-1) + m_j,
\end{equation}
for $j\geqslant 1$, where $m_j\in \{0, 1, \ldots, M-1\}$. Therefore, the shift parameter is
\begin{equation}
n(j) = \displaystyle \sum_{\ell = 1}^{j} \m_{\ell} M^{j-\ell} \in \{0, 1, \ldots, M^j-1\} 
\label{eqdpbis}
\end{equation}
at every resolution level $j$.
By construction, each $\wp_{j,n(j)}$ is obtained by recursively decomposing $\mathbf{U}$ \emph{via} a particular sequence of filters $(h_{\m_{\ell}})_{\ell=1,2, \ldots, j}$ where each $m_\ell$ belongs to $\{0, 1, \ldots, M-1\} $. 
Thus, path $\mathcal{P}$ can be associated with a unique $M$-ary sequence $\K=(\m_{\ell})_{\ell  \in \Nset}$ of elements of $\{0, 1, \ldots, M-1\}$. 
From now on, any given $M$-DWPT decomposition path will be represented by an $M$-ary sequence $\K$. 
Since the shift parameter $n$ depends on $j$ and $\K$ \emph{via} Eq. \eqref{eqdpbis}, the notation $n=n_{\K}(j)$ will hereafter be used to indicate this dependence if required. Therefore, an $M$-ary sequence $\K$ associated with an $M$-DWPT decomposition path specifies a unique sequence $(\wp_{j,n_{\K}(j)})_{j \in \Nset}$ of wavelet packets. 
Now we have:

\medskip
\begin{lemma}\label{lemma chemin approx}
Let $\mathcal{P}=\left(\mathbf{U}, \{\wp_{j,n_\K(j)}\}_{j \in \Nset}\right)$ be some path of the $M$-DWPT decomposition tree. 
If the shift parameter $n_\K$ is a bounded function of $j$, then $\K$ is the null sequence $\K_0=(0,0, \ldots)$. 
\end{lemma}

\medskip
\begin{IEEEproof} 
If there exists some $j_0$ such that $n_\K(j_0) \neq 0$, then it follows from Eq. \eqref{eqdpbis} that $n_\K(j)\geqslant M^{j-j_0}$ for every $j\geqslant j_0$.
Therefore, $n_\K$ cannot be upper-bounded by a constant. 
\end{IEEEproof}

\medskip
\begin{remark}\label{remark a propos de pesquet}
Lemma \ref{lemma chemin approx} states that the approximation path (associated with the null sequence) is the unique $M$-DWPT decomposition path for which the shift parameter $n$ is bounded by a constant independent of $j$. As a consequence, the results established in \cite[Corollary 5, Proposition 12]{pesquet99} and \cite[Proposition 7]{pesq07} concern the approximation path only, because the assumption that $n$ is constant is made in this reference to derive the asymptotic analysis. 
\end{remark}

\medskip
The results given in the present paper depends on the path chosen in the $M$-DWPT tree. In fact, the $M$-ary representation of the $M$-DWPT paths plays an important role throughout because it allows a complete path characterization. 
The following lemma will prove useful in the sequel.
\medskip
\begin{lemma}\label{dp}
For $n = n(j)$ given by Eq. \eqref{eqdpbis},
we have 
\begin{equation}
\F{\wpf}_{n}(\omega) = \left[ \prod_{\ell=1}^{j} H_{\m_{\ell}}(\frac{\omega}{M^{j+1-\ell}}) \right] \F{\Phi}(\frac{\omega}{M^{j}}). 
\label{eqlemma}
\end{equation}
\end{lemma}

\medskip
\begin{IEEEproof}
An easy extension of \cite[Lemma 1]{atto07}. 
\end{IEEEproof}

\subsection{Shannon $M$-DWPT and the Paley-Wiener space of $\pi$ band-limited functions}\label{Mband1.1}

The Shannon $M$-DWPT filters are hereafter denoted $h_{\m}^\S$ for $\m = 0, 1, \ldots, M-1$. These filters are ideal low-pass, band-pass and high-pass filters. We have
\begin{equation}
H_{\m}^\S(\omega) = \sum_{\ell \in \mathbb{Z}}  \un_{\Delta_{\m}}(\omega-2\pi \ell),
\label{HmShannon}
\end{equation}
where $\un_K$ denotes the indicator function of a given set $K$: $\un_K(x) = 1$ if $x \in K$ and $\un_K(x) = 0$, otherwise, and $\Delta_{\m}= \left[ -\frac{(\m+1)\pi}{M} , -\frac{\m\pi}{M} \right] \cup \left[\frac{\m\pi}{M} , \frac{(\m+1)\pi}{M} \right]$. 
The scaling function $\Phi^\S$ associated with these filters is defined for every $t \in \Rset$ by $\Phi^\S(t) = \sinc(t) = {\sin(\pi t)}/{\pi t}$ with $\Phi^\S(0) = 1$. The Fourier transform of this scaling function is
\begin{equation}
\F{\Phi^\S}=\un_{[-\pi,\pi]}.
\label{eqshanscal}
\end{equation}

The closure $\mathbf{U}^\S$ of the space spanned by the orthonormal system $\{ \tau_k \Phi^\S: k \in \Zset \}$ is then the Paley-Wiener (PW) space of those elements of $L^2(\mathbb{R})$ that are $\pi$ band-limited in the sense that their Fourier transform is supported within $[-\pi,\pi]$. 

Let $X$ be any band-limited Wide-Sense Stationary (WSS) random process whose spectrum is supported within $[-\pi, \pi]$. We have (see \cite[Appendix D]{atto07})
\begin{equation}
X[k]=\int_{\mathbb{R}} X(t)\Phi^\S(t-k) dt,
\label{eqechantillon}
\end{equation}
so that $\mathbf{U}^\S$ is the natural representation space of such a process. Any $M$-DWPT of $X$ can thus be initialized with the samples $X[k]$, $k \in \mathbb{Z}$.

Now, let us consider the Shannon $M$-DWPT of the PW space $\mathbf{U}^\S$. The wavelet packet functions $\wpf_{j,n}^\S$ of this $M$-DWPT can be computed by means of Eqs. \eqref{eqWPFwjn}, \eqref{eq winit} and \eqref{eqWPF-bis}, by setting $\Phi = \Phi^\S$ and $H_m = H_m^\S$, $m = 0, 1, \ldots, M-1$. The Fourier transforms of these wavelet packet functions are given by proposition \ref{propShannon} below, which extends \cite[Proposition 8.2, p. 328]{mallat99} since the latter follows from the former with $M = 2$.

\medskip
\begin{proposition}\label{propShannon}
For every non-negative integer $j$ and every $n \in \{0, \ldots, M^j-1\}$, 
\begin{equation}
\F{\wpf_{j,n}^\S} = M^{j/2} \un_{\Delta_{j,G(n)}},
\label{eq33modulobis}
\end{equation}
where, for any non-negative integer $k$,
\begin{equation}
\Delta_{j,k} = \left[ -\frac{(k+1)\pi}{M^{j}} , -\frac{k \pi}{M^{j}} \right]
\cup \left[\frac{k \pi}{M^{j}} , \frac{(k+1)\pi}{M^{j}} \right] 
\label{eq33modulo}
\end{equation}
and $G$ is defined by recursively setting, for $\m = 0, 1, \ldots, M-1$ and $\ell = 0, 1, 2, \ldots$, $G(0)=0$ and
\begin{equation}
G(M \ell + \m)=
\left\{
\begin{array}{l}
\begin{array}{l c c}
M G(\ell) + \m & \textrm{if} \: G(\ell) \: \textrm{is even},\\
M G(\ell) - \m + M-1 & \textrm{if} \: G(\ell) \: \textrm{is odd}.
\end{array}
\end{array}
\right.
\label{eqGpermute}
\end{equation} 
\end{proposition}

\medskip
In the rest of the paper, we set, for any pair $(j,k)$ of non-negative integers,
\begin{equation}
\Delta_{j,k}^{+} = \left[\frac{k \pi}{M^{j}} , \frac{(k+1)\pi}{M^{j}} \right].
\label{deltajp+}
\end{equation}

\section{Asymptotic analysis for the autocorrelation functions of the $M$-DWPT of second-order WSS random processes}\label{Section Decorrelation}

Let $X$ denote a centred second-order real random process assumed to be continuous in quadratic mean. The autocorrelation function of $X$, denoted by $\R$, is defined by $ \R(t,s) = \mathbb{E}[X(t) {X(s)} ]$.
The projection of $X$ on $\wp_{j,n}$ yields a sequence of random variables, the wavelet packet \emph{coefficients} of $X$:
\begin{equation}
c_{j,n}[k] = \displaystyle \int_{\mathbb{R}} X(t) \wpf_{j,n,k}(t) dt, \quad k \in \mathbb{Z},
\label{eq11}
\end{equation}
provided that the integral
\begin{equation}
\iint_{\mathbb{R}^2} \R(t,s) \wpf_{j,n,k}(t) \wpf_{j,n,k}(s) dt ds 
\end{equation}
exists, which will be assumed in the rest of the paper since commonly used wavelet functions are compactly supported or have sufficiently fast decay. The sequence given by Eq. \eqref{eq11} defines the discrete random process $c_{j,n}=(c_{j,n}[k])_{k\in\mathbb{Z}}$ of the wavelet packet \emph{coefficients} of $X$ at any resolution level $j$ and for any shift parameter $n\in \{0, 1, \ldots, M^j-1\}$.

\subsection{Problem formulation}\label{Assymp Decorrelation 1}

Let $R_{{j,n}}$ stand for the autocorrelation function of the random process $c_{j,n}$. We have
\begin{IEEEeqnarray}{l}
R_{{j,n}}[k,\ell] =\mathbb{E} \big [ c_{j,n}[k] {c_{j,n}[\ell]} \big ] 
=
\iint_{\mathbb{R}^2} \R(t,s) \wpf_{j,n,k}(t) \wpf_{j,n,\ell}(s) dt ds.
\label{eqRCJ} 
\end{IEEEeqnarray}

If $X$ is WSS, we write $\R(t,s) = \R(t-s)$ with some usual and slight abuse of language. From Eq. \eqref{eqRCJ}, it follows that 
\begin{equation}
R_{{j,n}}[k,\ell] = \iint_{\mathbb{R}^2} \R(t) \wpf_{j,n,k}(t+s) \wpf_{j,n,\ell}(s) dt ds.
\label{eq25renew}
\end{equation}

In the sequel, the spectrum $\gamma$ of $X$, that is, the Fourier transform of $\R$, is assumed to exist. 
By using Fubini's theorem and Parseval's equality, we can proceed as in \cite[Appendix C]{atto07} to derive from Eqs. \eqref{eqWPFjnk}, \eqref{eqWPFwjn} and \eqref{eq25renew} that $c_{j,n}$ is WSS. For any $k, \ell \in \mathbb{Z}$, and with the same abuse of language as above, the value $R_{{j,n}}[k,\ell]$ of the autocorrelation function of the discrete random process $c_{j,n}$ is $R_{{j,n}}[k-\ell]$ with
\begin{equation}
R_{{j,n}}[k] =  \frac{1}{2\pi} \int_{\mathbb{R}} \gamma(\frac{\omega}{M^j}) \vert \F{\wpf_{n}}(\omega) \vert^2  \exp \left ( {i k \omega} \right ) d\omega.
\label{eqauto}
\end{equation}

Let us assume that $\gamma \in L^{\infty}(\mathbb{R})$ and is continuous at $0$. These two assumptions have two easy consequences. First, the integrand on the right hand side (rhs) of Eq. \eqref{eqauto} is integrable since its absolute value is upper-bounded by $\Vert \gamma \Vert_{\infty}  \vert \F{\wpf_{n}}(\cdot) \vert^2$, whose integral equals $\Vert \gamma \Vert_{\infty}$; second, the limit of $\gamma(\frac{\omega}{M^j})$ is $\gamma(0)$ when $j$ tends to $\infty$. Therefore, for every given natural number $n$, it follows from Lebesgue's dominated convergence theorem applied to Eq. \eqref{eqauto} that
\begin{IEEEeqnarray}{rCl}
\displaystyle \lim_{j\rightarrow +\infty}  R_{{j,n}}[k] 
& = & \frac{1}{2\pi}  \displaystyle \int_{\mathbb{R}} \gamma(0) \vert \F{\wpf}_{n}(\omega) \vert^2  \exp \left ( {i k \omega} \right ) d\omega, \nonumber \\
& = & \gamma(0) \displaystyle \int_{\mathbb{R}} {\wpf}_{n}(t) {\wpf}_{n}(t-k)  dt = \gamma(0) \delta[k],
\label{eq rjn tend vers 0}
\end{IEEEeqnarray}
where $\delta$ is the standard Kronecker symbol defined for $k \in \mathbb{Z}$ by 
$
\delta[k]=
\left\{
\begin{array}{l}
\begin{array}{c c c}
1 & \textrm{if} & m=0,\\
0 & \textrm{if} & m\neq 0.
\end{array}
\end{array}
\right.
$

\noindent
The result thus obtained is that given in \cite[Corollary 5]{pesquet99}.


\medskip
From Lemma \ref{lemma chemin approx}, we distinguish two cases, for any given $M$-DWPT path $\mathcal{P}=\left(\mathbf{U}, \{\wp_{j,n_\K(j)}\}_{j \in \Nset}\right)$. 
First, if $n_\K$ is a constant function of $j$, then $\K$ is the null sequence $\K_0$, and thus, $\mathcal{P}$ is the approximation path. 
In this case, the shift parameter $n_\K(j)$ is $0$ at each resolution level $j$ and the $M$-DWPT of $X$ through path $\mathcal{P}=\mathcal{P}_{\K_0}$ consists of an infinite sequence of low-pass filters. The decorrelation is then guaranteed by Eq. \eqref{eq rjn tend vers 0} (see also \cite[Corollary 5]{pesquet99}). 
The second case is that of a function $n_\K$ which cannot be upper-bounded by a constant independent of $j$ when $j$ tends to infinity\footnote{Example: for the sequence $\K = (1, 1, \ldots)$, we have $n_\K(j) = M^j-1$ . The nodes $(j,M^j-1)$ are those of the path located at the extreme rhs of the $M$-DWPT decomposition tree.}.  In such cases where $n_\K$ is not a constant function of $j$, the asymptotic decorrelation of the $M$-DWPT coefficients at node $(j,n_\K(j))$ when $j$ tends to $\infty$ is no longer a mere consequence of Eq. \eqref{eq rjn tend vers 0} since Lebesgue's dominated convergence theorem does not apply. To proceed, we then write Eq. \eqref{eqauto} in the form
\begin{equation}
R_{j,n}[k] =  \frac{1}{2\pi} \int_{\mathbb{R}} \gamma(\omega) \vert \F{\wpf_{j, n}}(\omega) \vert^2  \exp \left ( {i M^j k \omega} \right ) d\omega.
\label{eq40ter}
\end{equation}
This equality derives from Eq. \eqref{eqauto} after a change of variable and by taking into account Eq. \eqref{eqWPFwjn}.

The purpose of the next section is then to analyse the behaviour of $R_{j,n}$ in the case of the Shannon filters and some families of filters that converge to the Shannon filters. 

\subsection{Asymptotic decorrelation achieved by the Shannon $M$-DWPT}\label{Assymp Decorrelation 2}


Let $\K = (\m_\ell)_{\ell \in \mathbb{N}}$ be an $M$-ary sequence of elements of $\{0, 1, \ldots, M-1\}$. Consider the Shannon $M$-DWPT, that is, the decomposition of $\mathbf{U^\S}$ associated with the Shannon $M$-DWPT filters $(h_{\m}^{\S})_{\m = 0, 1, \ldots, M-1}$. Let $\mathcal{P}_{\K} = (\mathbf{U^\S}, \{\wp_{j,n_{\K}(j)}^\S\}_{j \in \Nset})$ be the path associated with $\K$ in the Shannon $M$-DWPT decomposition tree. It follows from proposition \ref{propShannon} that the support of $\wpf_{j,n_{\K}(j)}^\S$ is $\Delta_{j,p_{\K}(j)}$, where $p_{\K}(j)=G(n_{\K}(j))$. For $j \in \Nset$, the sets $\Delta_{j,p_{\K}^+(j)}$ are nested closed intervals whose diameters tend to $0$. Therefore, their intersection contains only one point $\akappa$. It then follows from \eqref{eq33modulo} that
\begin{equation} 
\akappa = \lim_{j\rightarrow +\infty} \frac{p_{\K}(j)\pi}{M^j}.
\label{eq42}
\end{equation}

Let $X$ be some centred second-order WSS random process, continuous in quadratic mean, with spectrum $\gamma$. The autocorrelation function $R_{j,n}^\S$ resulting from the projection of $X$ on $\wp_{j,n}^\S$ derives from Eq. \eqref{eq40ter} and is given by
\begin{equation}
R_{j,n}^\S[k] = \frac{1}{2\pi} \int_{\mathbb{R}} \gamma(\omega) \vert \F{\wpf_{j,n}^{\S}}(\omega) \vert^2  \exp \left ( {i M^j k \omega} \right ) d\omega. 
\label{eq40-}
\end{equation}
From Eqs. \eqref{eq33modulobis} and \eqref{eq40-} and by taking into account that $\gamma$ is even, as the Fourier transform of the even function $\R$, it follows that
\begin{equation}
R_{j,n}^\S[k] 
=
\frac{M^j}{\pi} \displaystyle \int_{\Delta_{j,p}^{+}} \NNNN \gamma(\omega)\cos{(M^j k \omega)}  d\omega.
\label{eqproof1-}
\end{equation}
where $\Delta_{j,p}^{+}$ is given by Eq. \eqref{deltajp+} and $p = G(n)$. When $X$ satisfies some additional assumptions, the following theorem \ref{thmShannon} states that the Shannon $M$-DWPT of $X$ yields coefficients that tend to be decorrelated when $j$ tends to infinity. One of these additional assumptions is that $X$ is band-limited in the sense that its spectrum is supported within $[-\pi,\pi]$. When $M=2$, theorem \ref{thmShannon} is equivalent to \cite[Proposition 1]{atto07}. 

\medskip
\begin{theorem}\label{thmShannon}
Let $X$ be a centred second-order WSS random process, continuous in quadratic mean. 
Assume that the spectrum $\gamma$ of $X$ is an element of $L^{\infty}(\mathbb{R})$ and is supported within $[-\pi, \pi]$. Let $\K = (\m_\ell)_{\ell \in \mathbb{N}}$ be an $M$-ary sequence of elements of $\{0, 1, \ldots, M-1\}$
and $\mathcal{P}_{\K}=(\mathbf{U}^\S,\{\wp_{j,n_{\K}(j)}^\S\}_{j \in \Nset})$ be the Shannon $M$-DWPT decomposition path associated with $\K$. 

If the spectrum $\gamma$ of $X$ is continuous at point $\akappa$, then
\begin{equation} 
\lim_{j\rightarrow +\infty} \R_{j,n_{\K}(j)}^\S[k] = \gamma(\akappa)\delta[k]
\label{eq41}
\end{equation}
uniformly in $k \in \mathbb{Z}$, where $\R_{j,n_{\K}(j)}^\S$ is the autocorrelation function of the coefficients resulting from the projection of $X$ on $\wp^\S_{j,n_{\K}(j)}$.
\end{theorem}

\medskip
\begin{IEEEproof} 
The proof is an easy generalisation of that of \cite[Proposition 1]{atto07}, which concerns the standard wavelet packet transform ($M=2$). The key point of the proof is proposition \ref{propShannon} above, which makes it possible to compute Eq. \eqref{eq42}.
\end{IEEEproof}

\medskip
The foregoing theorem is mainly of theoretical interest since the Shannon $M$-DWPT filters have infinite supports and are not really suitable for practical purpose. In order to obtain a result of the same type for filters of practical interest, the $M$-DWPT is now assumed to be performed by using decomposition filters of order $\r$, $h_{\m}^{[\r]}$, $\m = 0, 1, \ldots, M-1$, such that
\begin{equation} 
\lim_{\r \rightarrow \infty} H_{\m}^{[\r]} =  H_{\m}^\S  \quad (\textrm{a.e.}).
\label{eq45}
\end{equation}
The parameter $r$ is called the order of the $M$-DWPT filters. 
According to \cite{shen98}, the Daubechies filters satisfy Eq. \eqref{eq45} for $M = 2$ when $\r$ is the number of vanishing moments of the Daubechies wavelet function; according to \cite{Aldroubi}, Battle-Lemari\'e filters also satisfy Eq. \eqref{eq45} for $M = 2$ when $\r$ is the spline order of the Battle-Lemari\'e scaling function. The existence of such families for $M > 2$ remains an open issue to address in forthcoming work. However, it seems reasonable to expect that general $M$-DWPT filters of the Daubechies or Battle-Lemari\'e type converge to the Shannon filters in the sense given above.

\medskip
\begin{theorem}\label{thm2}
Let $X$ be a centred second-order WSS random process, continuous in quadratic mean. 
Assume that the spectrum $\gamma$ of $X$ is an element of $L^{\infty}(\mathbb{R})$ and is supported within $[-\pi, \pi]$. Assume that the $M$-DWPT of the PW space $\mathbf{U}^\S$ is achieved by using decomposition filters $h_{\m}^{[\r]}$, $m = 0, 1, \ldots, M-1$, satisfying Eq. \eqref{eq45}. 

For every natural number $j$ and every $n = 0, 1, \ldots, M^j-1$, let $R_{{j,n}}^{[\r]}$ stand for the autocorrelation function of the wavelet packet coefficients of $X$ with respect to the packet $\wp^{[\r]}_{j,n}$.
We have
\begin{equation}
\lim_{r\rightarrow +\infty}
R_{{j,n}}^{[\r]}[k]= R_{{j,n}}^\S[k],
\label{eq47suite2}
\end{equation}
uniformly in $k \in \Zset$ and $n$, where $R_{{j,n}}^\S$ is given by Eq. \eqref{eqproof1-}.
\end{theorem}

\medskip
\begin{remark}
Since the $M$-DWPT concerns the space $\mathbf{U}^\S$, we have 
\begin{equation}
\F{\wpf_{n}^\S}(\omega) = \left [ \prod_{\ell=1}^{j} H_{\m_{\ell}}^\S(\frac{\omega}{M^{j+1-\ell}}) \right ] \F{\Phi^\S}(\frac{\omega}{M^{j}}),
\label{eqdp}
\end{equation}
\begin{equation}
\F{\wpf_{n}^{[\r]}}(\omega) = \left [ \prod_{\ell=1}^{j} H_{\m_{\ell}}^{[\r]}(\frac{\omega}{M^{j+1-\ell}}) \right ] \F{\Phi^\S}(\frac{\omega}{M^{j}}),
\label{eqdpblabla}
\end{equation}
where $n$ is given by Eq. \eqref{eqdpbis}. These equations
straightforwardly derive from Eq. \eqref{eqlemma} of lemma \ref{dp}. From Eqs. \eqref{eqWPFwjn}, \eqref{eq45}, \eqref{eqdp} and \eqref{eqdpblabla}, we obtain, for every given natural number $j$, that
\begin{equation}
\lim_{r\rightarrow +\infty}
\F{\wpf_{j,n}^{[\r]}} = \F{\wpf_{j,n}^\S} \quad (\textrm{a.e.}),
\label{eq47}
\end{equation}
uniformly in $n$. 
The three equalities above will prove useful below.
\end{remark}

\medskip
\begin{IEEEproof} \emph{(of theorem \ref{thm2})}. 
The autocorrelation function $\R_{j,n}^{[\r]}$ is given by Eq. \eqref{eq40ter} and is equal to
\begin{equation}
\R_{j,n}^{[\r]}[k] =  \frac{1}{2\pi} \int_{\mathbb{R}} \gamma(\omega) \vert \F{\wpf_{j,n}^{[\r]}}(\omega) \vert^2  \exp \left ( {i M^j k \omega} \right ) d\omega.
\label{eq46}
\end{equation}
In addition, we have
\begin{IEEEeqnarray}{l}
\big|\R_{j,n}^{[\r]}[k] - \R_{j,n}^{\S}[k] \: \big| 
\leqslant
\frac{1}{2\pi} \int_{\mathbb{R}} \vert \gamma(\omega) \vert \left| \:| \F{\wpf_{j,n}^{[\r]}}(\omega)|^2 - | \F{\wpf_{j,n}^\S}(\omega) |^2 \right| d\omega, 
\label{eqblabla}
\end{IEEEeqnarray}
where $\R_{j,n}^{\S}$ is given by Eq. \eqref{eqproof1-}. 
From Eqs. \eqref{eqWPFwjn}, \eqref{eqdp} and \eqref{eqdpblabla}, and by taking into acount that $|H_{\m_{\ell}}^{[\r]}(\omega)|$ and $|H_{\m_{\ell}}^\S(\omega)|$ are less than or equal to $1$ (due to the paraunitarity of the $M$-DWPT filters), we obtain 
\begin{equation}
\left| \:| \F{\wpf_{j,n}^{[\r]}}(\omega)|^2 - | \F{\wpf_{j,n}^\S}(\omega) |^2 \right| \leqslant 2 M^{j} \left|\F{\Phi^\S}(\omega) \right|^2.
\label{eq47suite1}
\end{equation}
The results derives from Eqs. \eqref{eq47}, \eqref{eqblabla}, \eqref{eq47suite1} and Lebesgue's dominated convergence theorem. 
\end{IEEEproof}

\section{Central limit theorems}\label{Loi}

We now consider a real random process $X$ that has finite cumulants and polyspectra. 
Denote by
\begin{IEEEeqnarray}{l}
{\Cum (t,s_1,s_2,\ldots,s_{\N})} 
=
\mathsf{cum} \{X(t), X(s_1), X(s_2), \ldots, X(s_{\N})\},
\end{IEEEeqnarray}
the cumulant of order $\N+1$ of $X$. The above cumulant is hereafter assumed to belong to $L^2(\mathbb{R}^{N+1})$ and to be finite for any natural number $\N$ (see \cite[Proposition 1]{pesq02} for a discussion about the existence of this cumulant). 
The cumulant of order $\N+1$ of the random process $c_{j,n}$ has the integral form given by (see \cite[Proposition 1]{pesq02}):
\begin{IEEEeqnarray}{l}
\Cum_{{j,n}} [ k,\ell_1, \ldots, \ell_{\N} ]  
= \mathsf{cum} \big \{ c_{j,n}[k] c_{j,n}[\ell_1] \ldots c_{j,n}[\ell_{\N}] \big \} \nonumber \\
\qquad \qquad = \int_{\mathbb{R}^{\N+1}} dt ds_1 \ldots ds_{\N} \Cum(t,s_1,s_2,\ldots,s_{\N}) \wpf_{j,n,k}(t) 
\wpf_{j,n,\ell_1}(s_1) \ldots \wpf_{j,n,\ell_{\N}}(s_{\N}).
\label{eqRCJcum1} 
\end{IEEEeqnarray}

Assume that $X$ is strictly stationary so that $\Cum(t,t+t_1,t+t_2,\ldots,t+t_{\N}) = \Cum(t_1,t_2,\ldots,t_{\N})$, then $c_{j,n}$ is a strictly stationary random process with cumulants $\Cum_{{j,n}}[k,k+k_1, k+k_2, \ldots, k+k_{\N}] = \Cum_{{j,n}}[k_1, k_2, \ldots, k_{\N}]$.
Assume also that $X$ has a polyspectrum $\gamma_N(\omega_1, \omega_2, \ldots, \omega_{\N})\in L^{\infty}(\Rset^N)$ for every natural number $N$ and every $(\omega_1, \omega_2, \ldots, \omega_{\N}) \in \Rset^N$. The polyspectrum is the Fourier transform of the cumulant $\Cum(t_1,t_2,\ldots,t_{\N})$. When $\N=1$, $\gamma_{1}$ is the spectrum of $X$ and is simply denoted $\gamma$ as in section \ref{Section Decorrelation}. Then after some routine algebra, Eq. \eqref{eqRCJcum1} reduces:
\begin{IEEEeqnarray}{l}
\Cum_{{j,n}}[k_1, \ldots, k_{\N}] =\frac{M^{-j({N-1})/{2}}}{(2\pi)^{\N}} \int_{\mathbb{R}^{\N}} 
d\omega_1 \ldots d\omega_{\N}  \exp \left ( {-i (k_1 \omega_1 + \ldots + k_{\N} \omega_{\N})} \right )   \nonumber \\
\qquad \cdots \gamma_{\N}(-{\omega_1}{M^{-j}}, \ldots, -{\omega_{\N}}{M^{-j}})    \F{\wpf_{n}}(-\omega_1 - \ldots - \omega_{\N})   \F{\wpf_{n}}(\omega_1)
\ldots
\F{\wpf_{n}}(\omega_{\N}).
\label{eqRCJcum4en0} 
\end{IEEEeqnarray}

If $n$ is a bounded function of $j$, it follows from Lebesgue's dominated convergence theorem that, for any natural number $N > 1$, $\Cum_{{j,n}}[k_1, k_2, \ldots, k_{\N}]$ tends to $0$ uniformly in $k_1, k_2, \ldots, k_N$ when $j$ tends to $\infty$. This is a consequence of \cite[Proposition 11]{pesquet99}. On the other hand, if $n$ cannot be upper-bounded by a constant independent with $j$, the situation is similar to that discussed in section \ref{Assymp Decorrelation 1}: the shift parameter $n$ depends on $j$ and Lebesgue's dominated convergence theorem does not apply to Eq. \eqref{eqRCJcum4en0} to prove the vanishing behaviour of the cumulants. The vanishing behaviour of the cumulants are hereafter given by theorems \ref{thm3} and \ref{thm4}. 

\medskip
By taking into account Eq. \eqref{eqWPFwjn}, Eq. \eqref{eqRCJcum4en0} can also be written, with an easy change of variables:
\begin{IEEEeqnarray}{l}
\Cum_{{j,n}}[k_1, \ldots, k_{\N}] =
\frac{1}{(2\pi)^{\N}}
\int_{\mathbb{R}^{\N}}
d\omega_1 \ldots d\omega_{\N} \exp \left ( {-i M^j (k_1 \omega_1 + \ldots + k_{\N} \omega_{\N})} \right )  \nonumber \\
\qquad \gamma_{\N}(-\omega_1, \ldots, -\omega_{\N})  \F{\wpf_{j, n}}(-\omega_1 - \ldots - \omega_{\N})  \F{\wpf_{j, n}}(\omega_1)  \ldots \F{\wpf_{j, n}}(\omega_{\N}).
\label{eqRCJcum4} 
\end{IEEEeqnarray}


\medskip
\begin{theorem}\label{thm3}
Let $X$ be a centred second-order strictly stationary random process, continuous in quadratic mean. Assume that the polyspectrum $\gamma_N$ of $X$ is an element of $L^{\infty}(\mathbb{R}^N)$ for any $N\geqslant 1$ and that the spectrum $\gamma$ is supported within $[-\pi, \pi]$. 
If $N>1$, then we have, uniformly in $k_1, k_2, \ldots, k_N$,
\begin{equation}
\lim_{j\rightarrow +\infty}
\Cum_{j,n}^{\S}[k_1, k_2, \ldots, k_N]= 0.
\label{eq48b}
\end{equation}
\end{theorem}

\medskip
\begin{IEEEproof}
When the wavelet packet functions are the functions $\wpf_{j,n}^{\S}$, it follows from Eqs. \eqref{eq33modulobis} and \eqref{eqRCJcum4} that the cumulant $\Cum_{{j,n}}^\S[k_1, k_2, \ldots, k_{\N}]$ of the discrete random process returned at node $(j,n)$ by the Shannon $M$-DWPT of $X$ satisfies
\begin{equation}
\vert \Cum_{{j,n}}^\S[k_1, k_2, \ldots, k_{\N}] \vert \leqslant 
\frac{M^{\frac{j(\N+1)}{2}} \Vert \gamma_{\N} \Vert_{\infty}}{(2\pi)^{\N}} 
\int_{\Delta_{j,p}^{\N}}
d\omega_1 d\omega_2 \ldots d\omega_{\N}
\label{eqProofThm3-1} 
\end{equation}
where $\Delta_{j,p}^{\N} = \underbrace{\Delta_{j,p} \times \Delta_{j,p} \times \ldots \times \Delta_{j,p}}_{N \hbox{ times}}$ and $p = G(n)$.

\medskip
According to Eq. \eqref{eq33modulo}, $\int_{\Delta_{j,p}} \! d\omega = 2\pi/M^j.$ Therefore, we obtain
\begin{equation}
\vert \Cum_{{j,n}}^\S[k_1, k_2, \ldots, k_{\N}] \vert \leqslant 
\Vert \gamma_{\N} \Vert_{\infty} M^{-j(\N-1)/2}.
\label{eqProofThm3-2} 
\end{equation}
Given any natural number $N > 1$, the rhs of the latter inequality does not depend on $n$, $k_1, \ldots, k_N$ and vanishes when $j$ tends to $\infty$, which completes the proof.
\end{IEEEproof}

\medskip
\begin{corollary}\label{thm4corrol}
With the same assumptions and notation as those of theorems \ref{thmShannon} and \ref{thm3}, assume that $\gamma$ is continuous at $\akappa$. Then, when $j$ tend to infinity, the sequence 
$\left(c_{j,n_{\K}(j)}^{\S}\right)_{\r, j}$ converges in the following distributional sense to a white Gaussian process with variance $\gamma(\akappa)$:
for every $x\in\mathbb{R}^\N$ and every $\epsilon>0$, there exists $j_0=j_0(x, \epsilon)>0$ such that, for every $j\geqslant j_0$, the absolute value of the difference between the value at $x$ of the probability distribution of the random vector $$(c_{j,n_{\K}(j)}^{\S}[k_1], c_{j,n_{\K}(j)}^{\S}[k_2], \ldots, c_{j,n_{\K}(j)}^{\S}[k_{\N}])$$ and the value at $x$ of the centred $\N$-variate normal distribution ${\cal N}(0, \gamma(\akappa) \I_{\N})$ with covariance matrix $\gamma(\akappa) \I_{\N}$ is less than $\epsilon$.
\end{corollary}

\medskip
\begin{IEEEproof}
A straightforward consequence of theorems \ref{thmShannon} and \ref{thm3}.
\end{IEEEproof}

\medskip
Consider filters satisfying Eq. \eqref{eq45}. Let $\K$ be an $M$-ary sequence of elements of $\{0, 1, \ldots, M-1\}$. 
The following results describe the asymptotic distribution of the discrete random process $c_{j,n_\K(j)}^{[r]}$ returned at node $(j, n_\K(j))$ when the resolution level $j$ and the order $r$ of the filters increase.

\medskip
\begin{theorem}\label{thm4}
Let $X$ be a centred second-order strictly stationary random process, continuous in quadratic mean. 
Assume that the polyspectrum $\gamma_N$ of $X$ is an element of $L^{\infty}(\mathbb{R}^N)$ for every natural number $N\geqslant 1$ and that the spectrum $\gamma$ is supported within $[-\pi, \pi]$. 

For every given natural number $j$ and every $n \in \{0, 1, \ldots, M^j-1 \}$, let $\Cum_{j,n}^{[\r]}$ stand for the cumulant of order $\N+1$ of the wavelet packet coefficients of $X$ with respect to the packet $\wp^{[\r]}_{j,n}$.

We have, uniformly in $n, k_1, k_2, \ldots, k_N$,
\begin{equation}
\hspace*{-0.5cm} \lim_{\r \rightarrow +\infty}
\Cum_{j,n}^{[\r]}[k_1, k_2, \ldots, k_N] 
= 
\Cum_{j,n}^{\S}[k_1, k_2, \ldots, k_N].
\label{eq48bcd}
\end{equation}
\end{theorem}

\medskip
\begin{IEEEproof}
By applying Eq. \eqref{eqRCJcum4} to $\wpf_{j,n}^{[r]}$ and $\wpf_{j,n}^{\S}$, we obtain
\begin{IEEEeqnarray}{l}
\vert \Cum_{{j,n}}^{[\r]}[k_1, \ldots, k_{\N}] - \Cum_{{j,n}}^{\S}[k_1, \ldots, k_{\N}] \vert  \leqslant
\frac{1}{(2\pi)^{\N}}
\vert \vert \gamma_{\N} \vert \vert_{\infty} 
\int_{\mathbb{R}^{\N}}
d\omega_1 \ldots d\omega_{\N}  \nonumber \\
\quad
\Big{|}
\F{\wpf_{j,n}^{[\r]}}(-\omega_1 \ldots - \omega_{\N}) 
\F{\wpf_{j,n}^{[\r]}}(\omega_1)
\ldots
\F{\wpf_{j,n}^{[\r]}}(\omega_{\N}) 
-
\F{\wpf_{j,n}^\S}(-\omega_1 \ldots - \omega_{\N}) 
\F{\wpf_{j,n}^\S}(\omega_1)
\ldots
\F{\wpf_{j,n}^\S}(\omega_{\N})
\Big{|}.  \nonumber
\\
\label{eqproof4-1} 
\end{IEEEeqnarray}
\noindent
The integrand on the rhs of the second inequality above can now be upper-bounded by 
\begin{IEEEeqnarray}{l}
2M^{j(\N+1)/2}\Phi^\S(\omega_1)\Phi^\S(\omega_2)\ldots \Phi^\S(\omega_{\N})
\label{eqproof4-1-1} 
\end{IEEEeqnarray}
where we use Eqs. \eqref{eqWPFwjn}, \eqref{eqdp}, \eqref{eqdpblabla}, and take into acount that $|H_{\m_{\ell}}^{[\r]}(\omega)|$ and $|H_{\m_{\ell}}^\S(\omega)|$ are less than or equal to $1$. 
The upper-bound given by Eq. \eqref{eqproof4-1-1} is independent of $\r$ and integrable; its integral equals $2M^{j(\N+1)/2}(2\pi)^N$. By taking Eq. \eqref{eq47} into account, we derive from Lebegue's dominated convergence theorem that the upper bound in Eq. \eqref{eqproof4-1} tends to $0$ when $\r$ tends to $+\infty$. 
\end{IEEEproof}

\medskip
\begin{corollary}\label{thm4corrol}
With the same assumptions and notations as those of theorems \ref{thm2} and \ref{thm4}, assume that $\gamma$ is continuous at $\akappa$. Then, when $j$ and $r$ tend to infinity, the sequence 
$\left(c_{j,n_{\K}(j)}^{[\r]}\right)_{\r, j}$ converges in the following distributional sense to a white Gaussian process with variance $\gamma(\akappa)$:
for every $x\in\mathbb{R}^\N$ and every $\epsilon>0$, there exists $j_0=j_0(x, \epsilon)>0$ and there exists $\r_0=\r_0(x, j_0, \epsilon)$ such that, for every $j\geqslant j_0$ and every $\r\geqslant \r_0$, the absolute value of the difference between the value at $x$ of the probability distribution of the random vector $$(c_{j,n_{\K}(j)}^{[\r]}[k_1], c_{j,n_{\K}(j)}^{[\r]}[k_2], \ldots, c_{j,n_{\K}(j)}^{[\r]}[k_{\N}])$$ and the value at $x$ of the centred $\N$-variate normal distribution ${\cal N}(0, \gamma(\akappa) \I_{\N})$ with covariance matrix $\gamma(\akappa) \I_{\N}$ is less than $\epsilon$.
\end{corollary}

\medskip
\begin{IEEEproof}
The result follows from Eqs. \eqref{eq41}, \eqref{eq47suite2}, \eqref{eq48b} and \eqref{eq48bcd}.
\end{IEEEproof}


\balance
\bibliographystyle{IEEEtran}
\bibliography{IEEEabrv,BibWAVELET}

\end{document}